\documentclass[useAMS,usenatbib,a4,referee]{mn2e}
\usepackage[dvips]{graphicx}

\title[The circumstellar environment of Wolf-Rayet stars \& Gamma-ray burst afterglows.]{The circumstellar environment of Wolf-Rayet stars \& Gamma-ray burst afterglows.}
\author[J. J. Eldridge, F. Genet, F. Daigne \& R. Mochkovitch]{J. J. Eldridge, F. Genet, F. Daigne \& R. Mochkovitch \thanks{E-mail: eldridge@iap.fr} \\ Institut d'Astrophysique de Paris, 98bis Boulevard Arago, 75014 Paris, France.\\ Universit\'e Pierre et Marie Curie -- Paris VI, 4 place Jussiey, 75005 Paris, France.}
\pagerange{\pageref{firstpage}--\pageref{lastpage}} \pubyear{2005}
\begin{document}
\maketitle
\label{firstpage}

\begin{abstract}
We study the evolution of the circumstellar medium of massive stars. We pay particular attention to Wolf-Rayet stars that are thought to be the progenitors of some long Gamma-Ray Bursts. We detail the mass-loss rates we use in our stellar evolution models and how we estimate the stellar wind speeds during different phases. With these details we simulate the interactions between the wind and the interstellar medium to predict the circumstellar environment around the stars at the time of core-collapse.
We then investigate how the structure of the environment might affect the GRB afterglow. We find that when the afterglow jet encounters the free-wind to stalled-wind interface that rebrightening occurs and a bump is seen in the afterglow light curve. However our predicted positions of this interface are too distant from the site of the GRB to reach while the afterglow remains observable.
\textbf{The values of the final-wind density, $A_{*}$, from our stellar models are of the same order ($\la 1$) as some of the values inferred from observed afterglow lightcurves. We do not reproduce the lowest $A_{*}$ values below 0.5 inferred from afterglow observations. For these cases we suggest that the progenitors could have been a WO type Wolf-Rayet star or a very low metallicity star.} Finally we turn our attention to the matter of stellar wind material producing absorption lines in the afterglow spectra. We discuss the observational signatures of two Wolf-Rayet stellar types, WC and WO, in the afterglow lightcurve and spectra. We also indicate how it may be possible to constrain the initial mass and metallicity of a GRB progenitor by using the inferred wind density and wind velocity.
\end{abstract}


\section{Introduction.}
Gamma-ray bursts (GRBs) are the most energetic and violent events in the Universe. While we understand the broad details of these events there is still much to learn. The observable details are limited to the initial flux of gamma-rays, the lightcurve of radio, optical and X-ray afterglows and the optical absorption line spectra of a few of these afterglows. The events themselves show individuality but there are similarities. We observe two types, long and short GRBs split by their duration. A GRB is long if the burst lasts for a time greater than two seconds and short otherwise. Also short bursts are observed to have a harder gamma-ray spectrum than their longer and softer counterparts. The two classes are thought to be explained by the same physics but differ in the progenitors.

Mystery still clouds short GRBs with the first X-ray afterglow only recently observed for GRB050509b \citep{aggrb050509b}. For long GRBs the first detection of an afterglow occurred for GRB970228 \citep{firstgrbag}. Observations of this and the following afterglows have increased our understanding of long GRBs. It is widely believed that the progenitors are massive stars that form a black hole at the end of their lives when nuclear reactions can no longer support the core against collapse. Then if the surrounding stellar material has enough angular momentum an accretion disk will form. This disk feeds the black hole that then produces a highly relativistic jet by a process that is not fully understood. If the star is small and compact the jet will emerge from the stellar surface and produce the GRB \citep{woosleyGRB,MW99,MWH01,zhanggrb}. After the initial event the jet continues its motion and is decelerated by the ambient medium  producing the afterglow.

However there are many exceptions to this standard model. There are some GRBs where the afterglow lightcurve rebrightens a few days after the burst. The proposed explanations are an inhomogeneous ambient medium, late energy input from the central engine or supernova (SN) light occurring from the same stellar death that gave rise to the GRB. The first example of a linked SN and GRB was GRB980425 and SN1998bw. However this event was peculiar with the GRB having an energy lower than a standard GRB and the SN being more energetic than a standard SN. The first definite relationship between a SN and GRB was when SN2003dh was discovered in the afterglow optical spectrum of GRB030329 \citep{stanek2003,hjorth2003}. 

The connection of SNe and GRBs adds weight to the argument for massive stars as the progenitors of GRBs. Therefore the evolution of massive stars has become vital in understanding long GRBs. Investigations have been split into two lines of enquiry. First has been the study of the progenitor stars themselves, for example \citet{izzy}, \citet{hmm2005}, \citet{langerthingy}, \citet{langeryoon}, \citet{hegerhemerge} and \citet{hegerlowZhomolog}. The preferred progenitors are Wolf-Rayet (WR) stars. They are massive stars that lose their hydrogen envelope to become naked helium stars. It has been suggested that at very low metallicities WR stars might be formed by fully homogeneous evolution, induced by rotation, during hydrogen burning \citep{langeryoon}. WR stars are preferred as they have a small radius so the relativistic jet can break out from the surface to produce the prompt emission. 

WR stars are expected to give rise to the hydrogen deficient type Ibc SNe. The number of these stars that also/otherwise produce a GRB at the end of their lives is important to know. The branching ratio of these events is uncertain but thought to be, $R({\rm GRB})/R({\rm SN Ibc})= 0.002$ to $0.004$ \citep{vanputten}. If this is true then something extra must occur to turn a SN progenitor into a GRB progenitor. These extra requirements for a GRB are that a black hole must form, probably directly, and the material close to the forming black hole must have enough angular momentum to form an accretion disk.

Studies with rotating models indicate that the core material of WR stars can retain enough angular momentum for a disk to be formed at the time of collapse \citep{hmm2005}. However the inclusion of magnetic fields introduces a mechanism to slow down the core, transferring angular momentum to the envelope. This means the core can no longer from a disk around the black hole \citep{langerthingy}. If the star is in a binary there are opportunities for the star to be spun up by mass transfer or tidal forces. This makes it more likely that the GRB progenitor occurs in a binary \citep{izzy,langerthingy}. Although models that include rotation, stellar magnetic fields and binary systems are all uncertain because they treat inherently three dimensional processes by one dimensional approximations. Also the observational evidence of the effects of rotation and interior magnetic fields on stellar evolution is mostly indirect.

In this study our models do not include rotation, stellar magnetic fields or binary stars. By using our models however we can take the first step to investigate the initial parameters of massive stars effect the circumstellar environment before we introduce these other complexities. Because of this our models will need something extra to turn them from SN progenitors to GRB progenitors. This something extra is likely to be non-solid body rotation or a binary companion. We will discuss the effect of these extra factors on our results.

The other theme of GRB progenitor studies concerns the circumstellar medium such as \citet{Enrico1}, \citet{Enrico2}, \citet{grbp2} and \citet{vanmarle}. These consider the environment around the stars through which the jet will propagate during its afterglow phase. This is a very important constraint since much more information can be gathered on the afterglow than can be obtained from the brief prompt gamma-ray emission. These studies are useful and indicate that some GRB afterglows are consistent with the circumstellar environment around WR stars. The two pieces of evidence are some afterglow lightcurves where the rate of decay is explained by the jet passing through a stellar wind environment. While in some GRBs absorption lines have been observed that are consistent with velocities from WR stars.

In this study we follow this line of enquiry, investigating the pre-SN winds and circumstellar environment around WR stars, to estimate the range of environments that can be expected around GRB progenitors if they are WR stars. We begin by briefly describing the most important details of our stellar model, the mass-loss prescription and the method used to calculate the velocity of the stellar wind during different phases of the stars' lifetimes. 

Using these details we then simulate the evolution of the circumstellar environment up to the point before a SNe or GRB will occur. We then discuss how the circumstellar environment varies over the stars initial mass and metallicity and the initial density of the interstellar medium (ISM). We pay particular attention to the free-wind density through which the jet will propagate through and whose density can be inferred from GRB afterglow lightcurves. 

From our simulations we suggest that in afterglow simulations a stalled-wind region should be included as well as the free-wind. The density jump between these two regions can produce rebrightening in the afterglow as suggested by \citet{grbsurf}. We demonstrate the effect on the lightcurve with our own afterglow calculations, varying the free-wind density and the position of the free-wind/stalled-wind interface.

Finally we turn to the observations of absorption features in afterglow spectra that may be caused by material in the GRB progenitor's stellar wind. First we consider the ionisation state of the material. Second we determine whether the velocities inferred from observations are consistent with our WR star models. We then show how using the absorption features and inferred wind density it is possible to constrain the initial mass and metallicity of GRB progenitors from afterglow lightcurves and optical absorption line spectra.

\section{Wolf-Rayet stars and their winds.}

The first GRB afterglow was discovered for GRB970228 \citep{firstgrbag}. At the time of writing 135 GRBs have observed optical afterglows\footnote{For an up to date list use the lists at: \\ \texttt{http://grad40.as.utexas.edu/tour.php} and \\ \texttt{http://www.mpe.mpg.de/$\sim$jcg/grbgen.html}.}. Many of these GRBs have detailed lightcurves and some have optical absorption line spectra. \citet{first4wrwinds} first noted that the absorption lines seen in the afterglow of GRB021004 where similar to the expected imprint of a wind from a WR star. Therefore it may be possible to directly measure the mass-loss rates and wind velocities of the massive stars that are GRB progenitors and compare the values to those predicted for stellar evolution models.

This requires the prediction of the range of wind velocities from massive stars and the environment around the star produced by the interaction of the wind with the interstellar medium (ISM). This information can be obtained by detailed stellar evolution calculations and consideration of hydrodynamics to estimate the density structure around the star. It is also important to estimate the ionisation state of this material to calculate which ionic species will be observed. The first step is to model the massive stars themselves.

\subsection{Construction and Testing of the Stellar Models.}

The stellar models we use were produced with the Cambridge STARS stellar evolution code originally developed by \citet{E71} and updated most recently by \citet{P95} and \citet{E03}. Further details can be found at the code's home page (http://www.ast.cam.ac.uk/$\sim$stars). The models are available from the same location for download. The models are the same as those described in \citet{EIT05}. We use 46 zero-age main-sequence models that have masses from $5$ to $200 \, {\rm M}_{\odot}$. The stars have a uniform composition determined by $X=0.75-2.5Z$ and $Y=0.25+1.5Z$, where $X$ is the mass fraction of hydrogen, $Y$ that of helium and $Z$ is the initial metallicity. This initial metallicity takes values from $10^{-3}$ to $0.05$, equivalent to $\frac{1}{20}Z_{\odot}$ to $2.5 Z_{\odot}$. The composition is taken to be scaled solar composition. 

All models have undergone carbon burning and most have started neon burning. The latest burning stages are very short, a standard WR wind of $1000 \, {\rm km \, s^{-1}}$ will only travel a distance of the order of ${\rm few} \times 10^{15}\, {\rm cm}$ as the final burning stages occur. This is smaller than the inner radius taken in our circumstellar environment simulations. 

The mass-loss prescription is based upon that of \citet{DT03} but has been modified. We use the rates of \citet{dJ} unless the mass-loss rates are described by one of the following prescriptions. For OB stars we use the rates of \citet{VKL2001} and for WR stars we use the rates of \citet{NL00}. The rates of \citet{VKL2001} include their own scaling with metallicity. For the remaining mass-loss rates we scale by the initial metallicity with the factor of $(Z/Z_{\odot})^{0.5}$. This is commonly used for non-WR stars \citep{ETsne,H03}, however it has only recently been suggested to be included for WR stars \citep{vanbevfull,WRZscale,Ethesis,EIT05}.

We must also note that while the error of the WR mass-loss rates quoted by \citet{NL00} is relatively small there is a larger possible systematic error in the mass-loss rates from the effect of clumping. The winds from hot stars are not smooth or uniform and tends to be clumped which higher and lower density clumps. This initially led to the WR mass-loss rates being overestimated by a factor of three. The problem of clumping in the winds of hot stars is still being tackled \citep{clump1,clump2} and therefore the magnitude of WR mass-loss remains uncertain by a factor of two.

\begin{figure}
\includegraphics[angle=0, width=84mm]{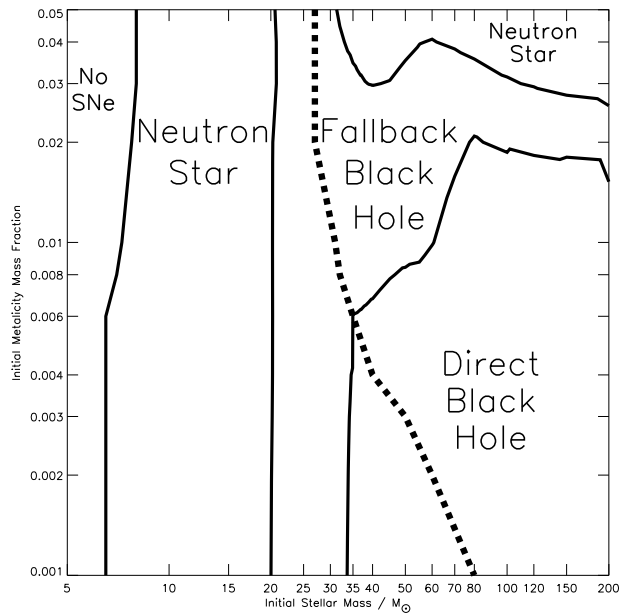}
\caption{Predicted compact remnant from stellar collapse. The solid lines separate regions where neutron stars, fallback black holes or direct black holes are formed. The dashed line is the minimum mass for Wolf-Rayet stars. To the left of this line the stars are Red Supergiants, to the right the stars are Wolf-Rayet Stars.}
\label{remnant}
\end{figure}

In figure \ref{remnant} we plot an important prediction of our models for a GRB progenitor. The solid lines represent the borders between the regions where different compact remnants are formed. Lower mass stars produce neutron star remnants. More massive stars produce black holes by fallback of material onto the forming neutron star. While the most massive stars have very massive cores that collapse directly to a black hole. We determine these regions using the method of \citet{H03} by examining the helium core mass. If the helium core mass is less than $8M_{\odot}$ a neutron star is formed, if the mass is greater than this a black hole is formed by fallback onto a neutron star. If the mass is greater than $15M_{\odot}$ a black hole is formed directly. 

The dashed line in figure \ref{remnant} is the minimum mass for hydrogen-free Wolf-Rayet stars. Stars to the left retain hydrogen while stars to the right have lost all hydrogen and are compact Wolf-Rayet stars. These are thought to be the most likely GRB progenitors as their compact size means the relativistic jet from the central black hole can penetrate through the stellar material and emerge at the surface to produce the GRB.

From the graph we notice that there is a maximum metallicity for direct black hole formation and that the minimum mass for a WR star increases with lower metallicity. These two factors are important in predicting the efficiency of producing GRB progenitors at different metallicities. If we require a GRB progenitor to be a WR star and form a black hole directly we can see the maximum number of stars meeting these two requirements is greatest at a metallicity mass fraction of 0.006 assuming the initial mass function and star-formation rate are constant with metallicity. At lower metallicities the number of stars decreases while at higher metallicities there is also a decrease, with a cut off at solar metallicity.

We note that all these stars are modelled as single stars and we do not include binary stars. We have investigated a small section of binary parameter space and find that the main result is to move the minimum mass for WR stars to lower initial masses. However the lowest initial mass for a binary WR star also increases with decreasing metallicity. Therefore there are fewer WR stars at lower metallicities and the binaries become the dominant channel for their production. At the lowest metallicities stars may evolve homogeneously becoming WR stars without losing a large amount of their initial mass \citep{langeryoon}.

We also do not include rotation or magnetic fields. The biggest effect of including rotation in massive stars is an enhanced mass-loss rate because of the stellar rotation \citep{hmm2004}, however there is little direct evidence that this should be included. The main effect on our models would be to produce a slightly lower final pre-SN mass. Rotating models tend to reproduce the WR stars rotating at velocities similar to those of observed WR stars \citep{wr_rotation}. However if the stars are rotating as solid bodies this is not fast enough rotation to produce a GRB. Therefore something extra is required for our models to be GRB progenitors. We discuss how rotation and duplicity are likely to affect our results.

\subsection{The Stellar Wind}

We have the mass-loss history for all our stellar models. We do not have the speed of the stellar wind as it is not required to model a star's evolution but it is required to model the circumstellar environment. There are few detailed studies of the dependence of wind velocity on stellar parameters. Fortunately the study of \citet{NL00} concerns the winds of WR stars. Therefore we can predict the density structure that surrounds these stars with reasonable accuracy.

\begin{table}
\caption{Value of $\beta_{\rm W}$ used for pre-WR stars in wind speed calculations.}
\label{windtable}
\begin{tabular}{lc}
\hline
$T_{\rm eff} / K$	&	$\beta_{\rm W}$\\
\hline
$<3600$			&	0.125\\
6000			& 	0.5\\
8000			&0.7\\
10000			&1.3\\
20000			&1.3\\
$>22000$			&2.6\\
\hline
\end{tabular}
\end{table} 

The first step in estimating stellar wind speeds is to determine the stellar type. We define two types, WR stars and pre-WR stars. We use the same definitions as for our mass-loss rates. If the surface hydrogen mass fraction is less than 0.4 and the surface temperature is greater than $10^{4}{\rm K}$ the star is a WR star. Other stars feature a large hydrogen envelope and are main-sequence stars or red supergiants (RSGs). We refer to these as pre-WR stars.

For pre-WR stars we first calculate the escape velocity at the stellar surface. It is a reasonable assumption that any material escaping from a star's gravitational influence must be related to this velocity. To account for the reduction of the escape velocity of luminous stars which are close to the Eddington Limit we include the Eddington factor. Therefore,
\begin{equation}
v_{\rm escape}^{2} = 2GM_{*}(1-\Gamma)/R_{*},
\end{equation}
when $\Gamma = L_{*}/L_{\rm Edd} = 7.66 \times 10^{-5} \sigma_{e} (L_{*}/M_{*})$, and $\sigma_{e}=(0.401 (X+Y/2+Y/4)) {\rm cm^{2}}$. The escape velocity without the Eddington factor, was taken directly as the wind velocity in the study by \citet{Enrico1}; this overestimates the wind speed. Further more the physical processes that accelerate the stellar wind are complex \citep{K87,VKL2001}. To account for this we relate the wind velocity and the escape velocity by using a constant to express the acceleration process uncertainty,
\begin{equation}
v_{\rm wind}^{2} = \beta_{\rm W} v_{\rm escape}^{2}.
\end{equation}

For OB stars we use the $\beta_{\rm W}$ from \citet{VKL2001} and supplement this with the values in \citet{HPT02}. We list the values used for $\beta_{W}$ in table \ref{windtable}. To obtain $\beta_{\rm W}$ between the ranges listed we use linear interpolation. The wind speeds calculated are very uncertain. There is an additional uncertainty in how changing metallicity affects these wind speeds. Our resultant estimated wind speeds are probably correct to an order of magnitude.

WR stellar wind velocities are calculated in a similar method. We use the empirical formulae of \citet{NL00}. WR stars have unique atmospheres, their wind is optically thick and not in hydrostatic or local thermodynamic equilibrium. Because of this, models of WR stars predict smaller stellar radii and higher surface temperatures than for observed WR stars. \citet{NL00} take this disparity into consideration and provide formulae that fit $\beta_{\rm W}$ to the luminosity and surface composition of WR stars.

There are two wind velocity formulae, one for WN stars and another for WC stars. The second letter indicates which element is most prominent in the WR star spectrum. N indicates nitrogen and C indicates carbon. The progression indicates the gradual exposure of nuclear burning products on the surface of the star. The last in the sequence are WO stars where oxygen is the most dominant element in the spectrum. Very few WO stars have been observed, their mass-loss rates agree with those of WC stars, however their wind velocities can be much higher than WC stars of equivalent luminosity. We assume WR stars to be WN stars unless all hydrogen has been removed and $(x_{C}+x_{O})/y>0.03$ when we have a WC, where $x_{C}$, $x_{O}$ and $y$ are the surface number fractions of carbon, oxygen and helium.

The wind velocity of a WN star, as derived by \citet{NL00} is,
\begin{equation}
\log(v_{\rm wind}/v_{\rm escape}) = 0.61-0.13 \log L_{*} + 0.30 \log Y,
\end{equation}
with a standard deviation of 0.084 dex. For a WC star the wind velocity is given by,
\begin{equation}
\log(v_{\rm wind}/v_{\rm escape}) = -2.37+0.13 \log L_{*} -0.07 \log Z.
\end{equation}
with a standard deviation of 0.13 dex. Using these equations gives a prediction of the wind velocity for our WR stars. Although as mentioned about there is an uncertainty in greater than those listed due to clumping of the WR wind.

\begin{figure}
\includegraphics[angle=0, width=84mm]{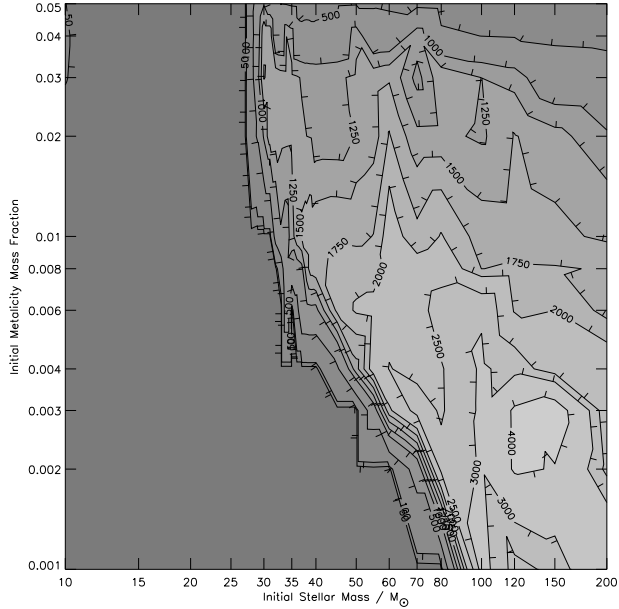}
\caption{Final wind velocity from the stellar models. Contour values are in $\rm km \, s^{-1}$.}
\label{velplot}
\end{figure}

In our wind speeds we include a factor of $(Z/Z_{\odot})^{0.13}$ as suggested by \citet{VKL2001}. In this we are assuming that all stellar winds are driven by radiative processes on metals. This is true for OB main sequence stars but the processes that drives WR and RSG mass loss are unknown even though there are some suggestions \citep{WRwinds1}. For WR stars it is more uncertain whether we should include the scaling of wind velocity by initial metallicity. In radiatively driven winds it is known that the few but strong lines of CNO elements determine the wind speed while the many but weak lines of iron group elements determine the mass-loss rate \citep{VKL2001}. The CNO abundance in WR stars has little relation to the initial metallicity. However we do include the metallicity scaling in calculating our predicted WR wind velocities. 

In figure \ref{velplot} we plot the pre-SN wind velocity of our models. During the evolution of the star the wind velocity varies above or below the values in the figure. The wind velocities indicated here will determine the free-wind region velocity and density. For single stars we can see there is a clear division between the RSGs and WR stars with them having slow and fast winds respectively. The evolution of pre-SN velocity with metallicity is interesting. WR stars at lower metallicity are more massive for the same initial mass due to reduced mass loss. This makes the escape velocity larger and therefore from our calculations the wind velocity greater. A second contributing factor is that more massive WR stars are more luminous, this too leads to stronger, faster wind. The wind velocities from the WN and WC star models, by construction, agree with the observed wind velocities of WN and WC stars listed in \citet{NL00}.

It is important to note that our models only produce WN or WC stars. We find no WO stars, if we adopt the normal \textit{theorist} definition of $(x_{C}+x_{O})/y>1$ \citep{sm1991,DT03}. \citet{woobs} present observations of five WO stars, one each in the SMC and LMC and three in the galaxy. Their wind velocities are in the range of 4200 to 5500 ${\rm km \, s^{-1}}$. Comparing these to other WR stars we find their their wind velocities are significantly larger by a factor of two to three than WC stars of the same luminosity at solar metallicity. Because there are very few observed WO stars it is not possible to determine a trend of how the wind velocity changes with the stellar parameters.

WO are very rare in the sky. They are also very difficult to produce in stellar models, \citet{hmm2005} produce one WO star from their collection of models. This is because the do not scale the WR mass-loss rates with metallicity. If we remove the scaling we also find a few WO stars at sub-solar metallicities. This however goes against the growing evidence for including the scaling discussed above. The other solution would be to change the definition of WO stars surface abundance, for example to $(x_{C}+x_{O})/y>0.8$. The observational evidence is not clear. \citet{woobs} observed one WO star with $(x_{C}+x_{O})/y=1.1$, for three other WO stars they found that the value to be 0.62. More recently \citet{woobs3} studied one of these objects and found a larger value of $0.85 \pm 0.2$. Other studies such as \citet{woobs2} and \citet{woobs4} also find higher values of 0.92 and 1.08 for other WO stars. The uncertainties in observationally inferred abundances make it difficult to draw any firm conclusion. We can conclude that the number of WO stars we find is sensitive to our definition of when a stellar model becomes a WO star.

Further more, as already noted, the WR mass-loss rates are uncertain. At solar metallicity if they were increased within the uncertainities we would find some WO stars. Whilst at lower metallicities, due to the scaling of mass loss with metallicity, it would still be difficult to produce WO stars. Alternatively WO stars do not occur for single, non-rotating stars and are the result of increased mass-loss during evolution, either by rotation or a binary interaction.

To account for our lack of WO stars in our models it must be remembered that supplemental to the wind velocities shown in figure \ref{velplot}, WO stars also produce wind velocities greater than $4000 {\rm km \, s^{-1}}$, over a range of metallicities. Therefore the velocities are uncertain and values up to 3 times the values presented here may occur. Further more we find that our solar metallicity WR models have higher wind speeds earlier in their WC evolution before slowing to the values in figure \ref{velplot}. This means that there are WC stars that have slightly higher velocities than our predicted pre-SN wind velocities.

There is a further uncertainty introduced into the final wind velocities from binary stars. Binary systems not only enhance mass-loss they introduce the possibility of mass gain, either my stellar mergers of mass transfer. If a star is replenished with hydrogen the helium core will be able to grow larger as the gained mass is lost. Larger helium cores will lead to more massive WR stars. More massive WR stars may also be possible through mergers of two helium stars as suggested by \citet{hegerhemerge}. Such stars will be more luminous and therefore will have faster winds with speeds intermediate between those in figure \ref{velplot} and those of WO stars.

Our next step is to simulate the interactions of the wind with the surrounding ISM. This will then enable us to predict the circumstellar environment in which the GRB jet propagates through and interacts with to produce a GRB afterglow.

\subsection{The circumstellar material}

The theory of stellar wind bubbles has long be established \citep{windbubbles1,windbubbles2,blastwaves}. These analytic studies provide the details of the interaction between stellar winds and the surrounding inter-stellar medium. How the various wind phases interact has been studied by \citet{garcia1}, \citet{garcia2}, \citet{Enrico2} and \citet{vanmarle}. We perform a similar study using the Zeus-2D code as described by \citet{Zeus}. An important difference is that we investigate how initial stellar mass, initial metallicity and initial ISM density effect the resultant wind bubbles.

\begin{figure}
\includegraphics[angle=0, width=84mm]{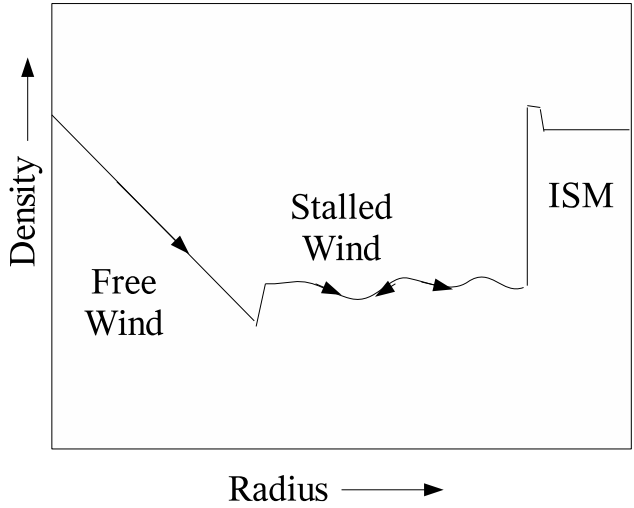}
\caption{A representative density profile of stellar wind bubbles. The axes units are plotted on a logarithmic scale.}
\label{simplefig}
\end{figure}

\begin{figure}
\includegraphics[angle=0, width=84mm]{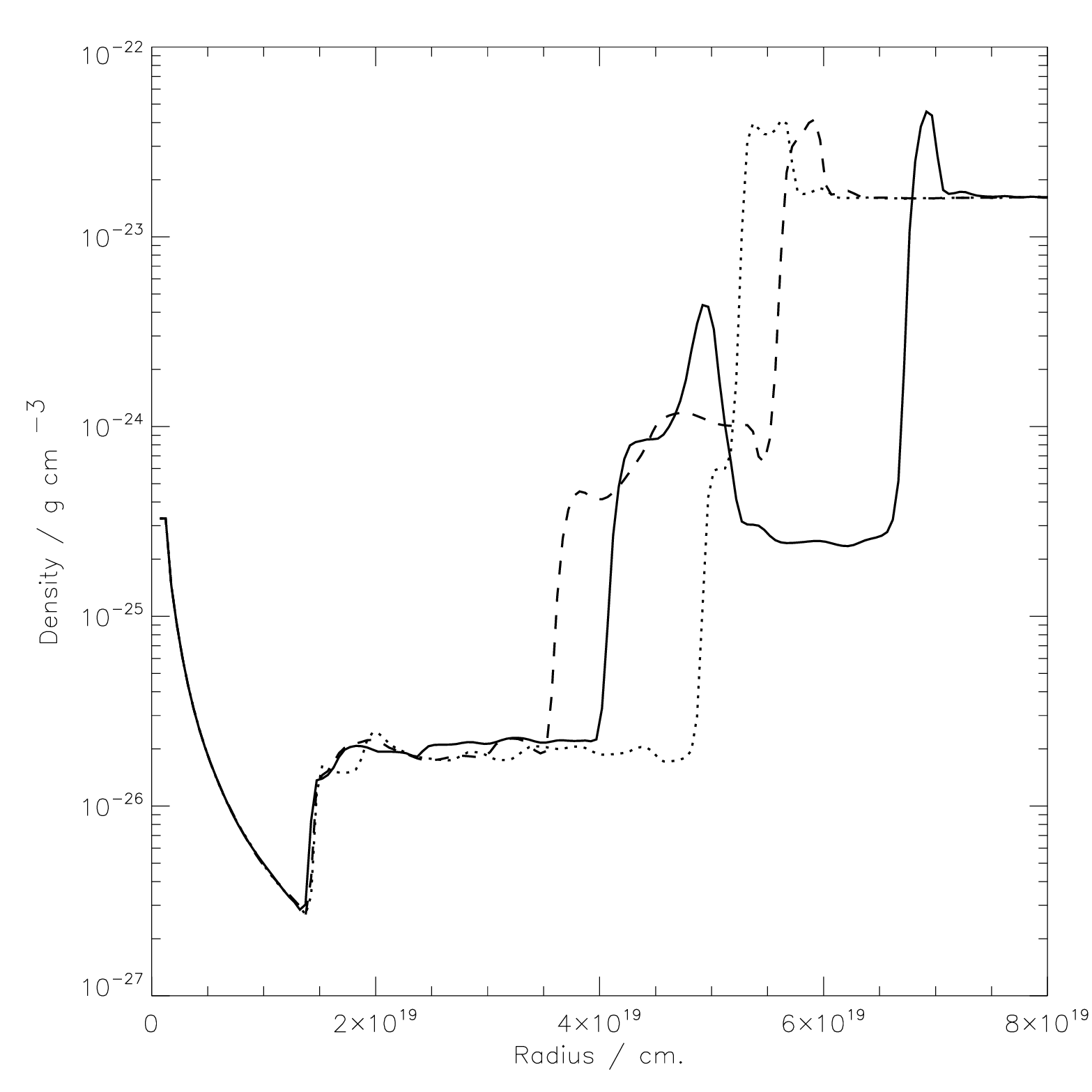}
\includegraphics[angle=0, width=84mm]{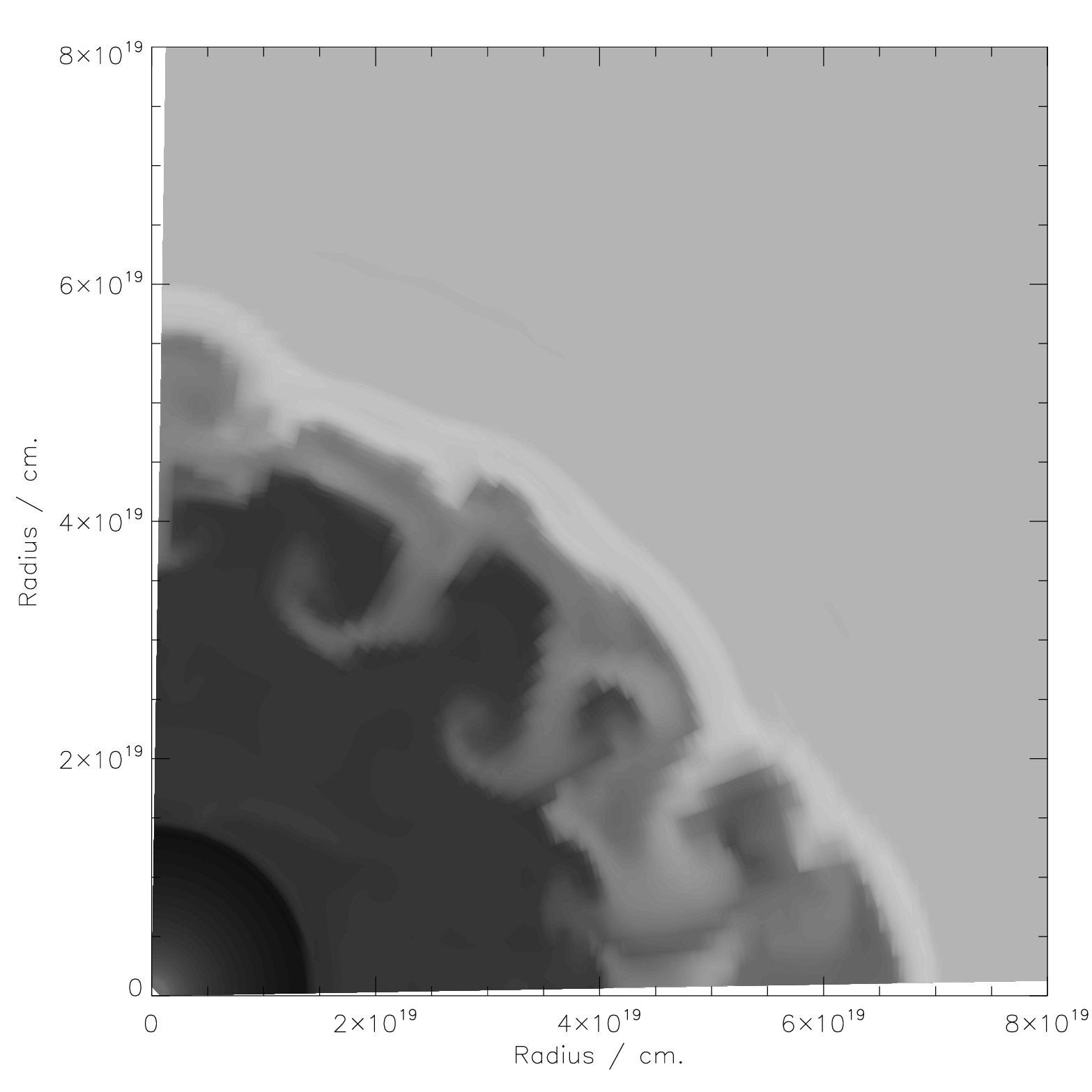}
\caption{The density of the circumstellar environment of a $70M_{\odot}$ star with $Z=0.004$. The initial ISM density was $n_{0}=10 \, {\rm cm^{-3}}$. The age of the star is 3.86 million years and the nuclear reaction in the core is neon burning. In the lower panel darker colours are less dense, lighter shades are higher density. In the upper panel we present cross-sections through this environment. The solid line follows the x-axis, the dotted line along the line $x=y$ and the dashed line the y-axis.}
\label{hydro2d}
\end{figure}

\begin{figure}
\includegraphics[angle=0, width=84mm]{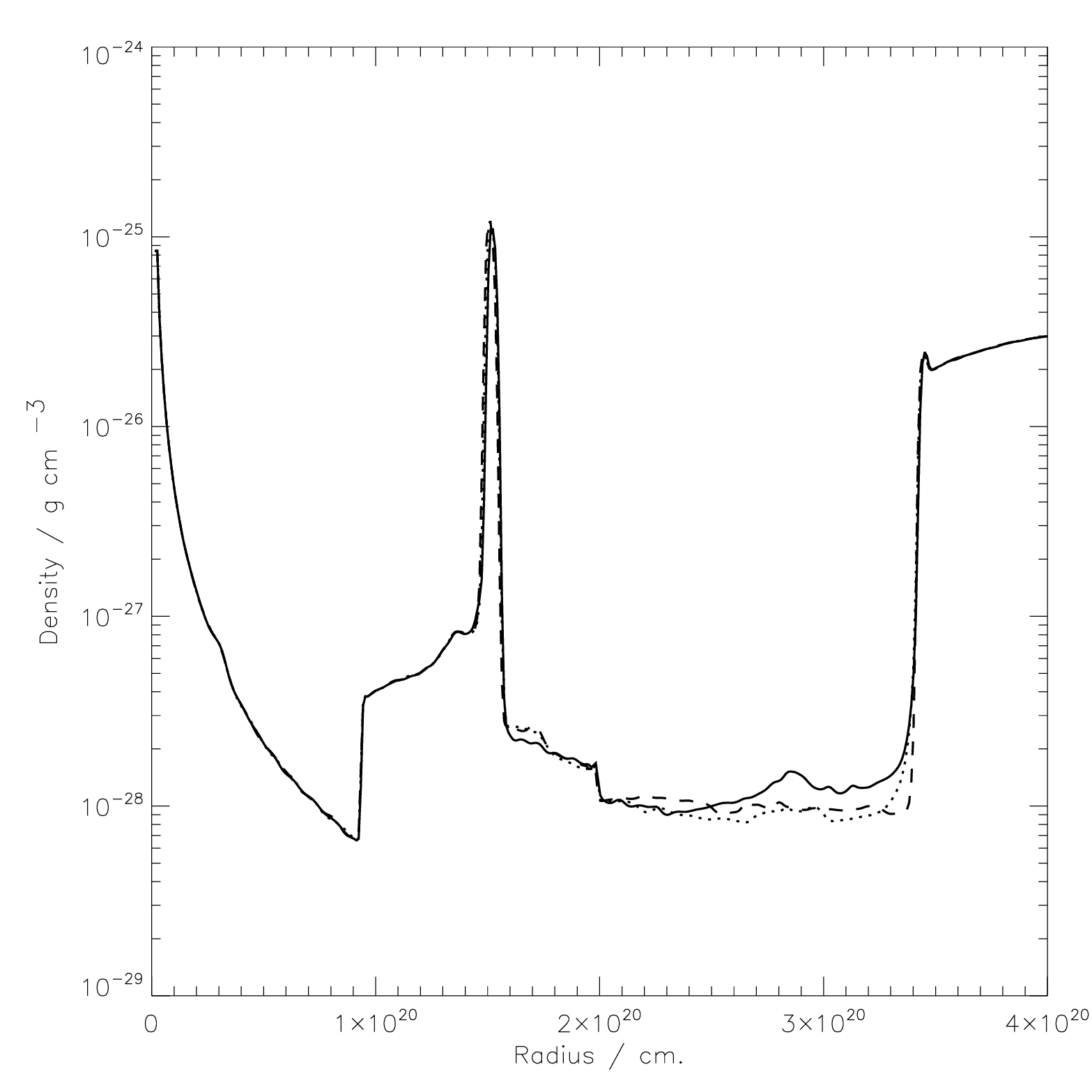}
\includegraphics[angle=0, width=84mm]{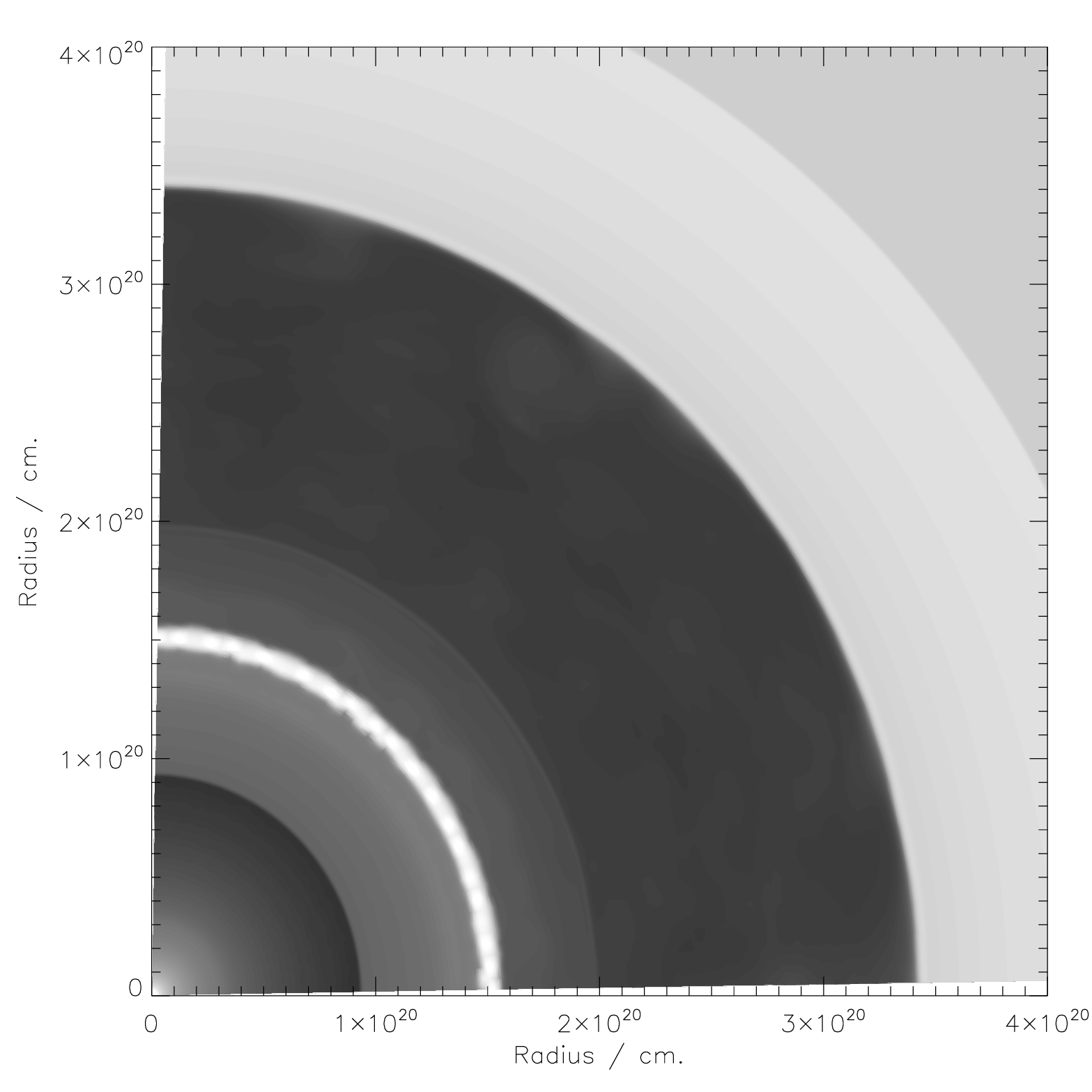}
\caption{This figure is the same as figure \ref{hydro2d} but the initial ISM density was $n_{0}=0.01 \,{\rm cm^{-3}}$.}
\label{hydro2d_B}
\end{figure}

Our simulations are simple. The simulations include only the hydrodynamics of the stellar wind and ISM, conserving momentum, mass and energy. We assume the material composition is homogeneous and introduce the wind material at the inner meshpoints at the correct density and velocity to give the mass-loss rate predicted from our stellar models. We allow the velocity of wind and the mass-loss rate to vary exactly as that predicted from the stellar models rather than chose a fixed average wind speed and mass-loss rate. We use 1000 meshpoints in the radial direction and 49 in the azimuthal direction over $90^{\circ}$. We first run 1D simulations to determine the radial extent required. We then rerun the simulations in 2D. The position of the outer meshpoint is varied from $5 \times 10^{19}$ to $10^{21} {\rm cm}$ and the inner mesh point varies from $5 \times 10^{16}$ to $10^{18} {\rm cm}$.

There are many factors that we have not included, that would make the simulations more exact. For example radiative cooling, thermal conduction, inhomogeneities in the ISM, magnetic fields, turbulence, inhomogeneity of the stellar wind, varying composition and radiative heating and ionisation. Including these details would alter the simulation results. For example \citet{garcia3} investigated the effect of including photoionisation on the size of the stellar wind bubble. They found the inclusion does affect the size of the bubble. However including all these details would slow down our code. Furthermore most are not well understood or constrained. Using simple models allows us to estimate the general shape of the environment we can expect and the relevant important distances within the circumstellar material to order of magnitude accuracy.

The evolution of the bubble begins during the main-sequence with a weak ($\approx 10^{-6} M_{\odot} {\rm yr}^{-1}$) and fast ($\approx 1000 \,{\rm km \, s}^{-1}$) wind. This creates the bubble in the ISM and is important for the overall size of the bubble as this wind endures the longest time. With higher ISM densities the size of this bubble can be very small as the wind is not powerful enough to push out against a dense ISM. Almost immediately after the bubble starts forming, it grows and organises itself into the structure shown in figure \ref{simplefig} with an inner free-wind region separated from a slowly expanding stalled-wind region by a shock; there is then a thin dense shell of swept-up and shocked ISM and then outermost the undisturbed ISM. There is a contact discontinuity between the ISM and wind material The flatness of the stalled-wind region depends on the initial ISM density and the inhomogeneity of the wind. We introduce random variation in the wind speed of the order of 1 percent over the azimuthal direction as in \citet{garcia1}. Increasing the magnitude of this factor results in more mixing, instabilities and inhomogeneities in the wind.

\begin{table*}
\begin{minipage}{160mm}
\caption{Distance of free-wind to stalled-wind interface. $n_{0}$ is the initial ISM density in ${\rm particles}/{\rm cm^{3}}$. $R_{\rm SW}$ is the radius of the free-wind/stalled-wind interface in $\rm cm$.}
\label{innerR}
\begin{tabular}{@{}cccccccccc@{}}
\hline

   $\log(n_{0}$&$ / {\rm cm^{-3}})$& & 3 & 2 & 1 & 0 & -1 & -2 & -3\\
\hline
$Z$ & $M/M_{\odot}$ & $A_{*}$   & $\log(R_{\rm SW})$ & $\log(R_{\rm SW})$ & $\log(R_{\rm SW})$ & $\log(R_{\rm SW})$& $\log(R_{\rm SW})$ & $\log(R_{\rm SW})$ &  $\log(R_{\rm SW})$\\
\hline
0.020 & 200& 0.93& 18.31&  18.88&  19.20&  19.61&  20.05&  20.26& 20.82\\
0.020 & 150& 0.76& 18.35&  18.86&  19.21&  19.52&  19.91&  20.50& 20.71\\
0.020 & 120& 0.75& 18.37&  18.85&  19.18&  19.55&  19.94&  20.54& 20.70\\
0.020 & 100& 1.06& 18.31&  18.81&  19.18&  19.62&  19.86&  20.52& 20.67\\
0.020 & 80 & 3.78& 18.39&  18.91&  19.28&  19.77&  20.21&  20.46& 20.30\\
0.020 & 70 & 3.14& 18.31&  18.83&  19.24&  19.62&  20.25&  20.43& 20.31\\
0.020 & 60 & 0.66& 18.38&  18.78&  19.12&  19.47&  19.80&  20.47& 20.57\\
0.020 & 50 & 0.78& 18.47&  18.74&  19.09&  19.48&  19.75&  20.47& 20.51\\
0.020 & 40 & 0.89& 18.31&  18.73&  19.09&  19.36&  19.86&  20.36& 20.39\\
0.020 & 30 & 2.61& 18.31&  18.77&  19.07&  19.56&  19.70&  19.81& 19.81\\
0.008 & 200& 4.41& 18.51&  18.94&  19.38&  19.82&  20.25&  20.31& 20.41\\
0.008 & 150& 3.90& 18.51&  18.90&  19.35&  19.66&  20.28&  20.40& 20.42\\
0.008 & 120& 3.09& 18.48&  18.89&  19.33&  19.66&  20.22&  20.32& 20.36\\
0.008 & 100& 5.52& 18.45&  18.85&  19.24&  19.70&  19.98&  20.43& 20.47\\
0.008 & 80 & 0.80& 18.48&  18.80&  19.21&  19.68&  19.98&  20.39& 20.41\\
0.008 & 70 & 0.74& 18.94&  18.79&  19.16&  19.62&  20.01&  20.35& 20.34\\
0.008 & 60 & 0.69& 18.43&  18.79&  19.15&  19.62&  20.14&  20.27& 20.28\\
0.008 & 50 & 0.68& 18.41&  18.76&  19.12&  19.45&  20.04&  20.09& 20.09\\
0.008 & 40 & 1.43& 18.35&  18.83&  19.21&  19.69&  19.82&  19.87& 19.87\\
0.004 & 200& 2.36& 18.58&  18.95&  19.43&  19.80&  20.18&  20.23& 20.23\\
0.004 & 150& 2.09& 18.59&  18.98&  19.34&  19.73&  20.14&  20.16& 20.23\\
0.004 & 120& 1.03& 18.56&  18.86&  19.23&  19.71&  20.10&  20.30& 20.36\\
0.004 & 100& 0.80& 18.50&  18.86&  19.19&  19.66&  20.17&  20.29& 20.29\\
0.004 & 80 & 0.85& 18.48&  18.85&  19.21&  19.56&  20.10&  20.16& 20.16\\
0.004 & 70 & 0.70& 18.40&  18.80&  19.15&  19.62&  19.91&  19.97& 19.97\\
0.004 & 60 & 0.55& 18.27&  18.76&  19.21&  19.62&  19.78&  19.78& 19.78\\
0.004 & 50 & 3.01& 18.40&  18.81&  19.20&  19.59&  19.81&  19.81& 19.81\\
0.001 & 200& 1.29& 18.56&  18.93&  19.36&  19.69&  19.70&  19.70& 19.70\\
0.001 & 150& 1.42& 18.45&  18.85&  19.32&  19.62&  19.72&  19.72& 19.72\\
0.001 & 120& 1.11& 18.30&  18.86&  19.37&  19.42&  19.42&  19.42& 19.42\\
0.001 & 100& 1.10& 18.39&  18.85&  19.25&  19.62&  19.75&  19.75& 19.75\\

\hline
\end{tabular}
\end{minipage}
\end{table*} 

\begin{table*}
\begin{minipage}{160mm}
\caption{Distance of stalled-wind to SW interface. $n_{0}$ is the initial ISM density in ${\rm particles}/{\rm cm^{3}}$. $R_{\rm ISM}$ is the radius of the stalled-wind/ISM interface in $\rm cm$. }
\label{outerR}
\begin{tabular}{@{}cccccccccc@{}}
\hline
   $\log(n_{0}$&$ / {\rm cm^{-3}})$& & 3 & 2 & 1 & 0 & -1 & -2 & -3\\
\hline
$Z$ & $M/M_{\odot}$ & $A_{*}$   & $\log(R_{\rm ISM})$ & $\log(R_{\rm ISM})$ & $\log(R_{\rm ISM})$ & $\log(R_{\rm ISM})$& $\log(R_{\rm ISM})$ & $\log(R_{\rm ISM})$ &  $\log(R_{\rm ISM})$\\
\hline

0.020 & 200& 0.93& 19.42& 19.78&  20.05& 20.32& 20.54& 20.68& 20.94\\
0.020 & 150& 0.76& 19.32& 19.75&  20.02& 20.29& 20.52& 20.74& 20.91\\
0.020 & 120& 0.75& 19.26& 19.69&  19.99& 20.27& 20.50& 20.72& 20.88\\
0.020 & 100& 1.06& 19.31& 19.58&  19.96& 20.24& 20.48& 20.70& 20.89\\
0.020 & 80 & 3.78& 19.19& 19.62&  19.90& 20.19& 20.44& 20.66& 20.87\\
0.020 & 70 & 3.14& 19.18& 19.59&  19.87& 20.17& 20.43& 20.65& 20.86\\
0.020 & 60 & 0.66& 19.23& 19.57&  19.87& 20.16& 20.36& 20.64& 20.84\\
0.020 & 50 & 0.78& 17.76& 19.51&  19.82& 20.12& 20.31& 20.62& 20.83\\
0.020 & 40 & 0.89& 19.02& 19.35&  19.75& 20.05& 20.33& 20.57& 20.79\\
0.020 & 30 & 2.61& 18.89& 19.20&  19.39& 19.89& 20.21& 20.50& 20.75\\
0.008 & 200& 4.41& 19.34& 19.69&  19.99& 20.25& 20.48& 20.68& 20.87\\
0.008 & 150& 3.90& 19.35& 19.65&  19.95& 20.19& 20.45& 20.65& 20.85\\
0.008 & 120& 3.09& 19.30& 19.60&  19.91& 20.19& 20.42& 20.64& 20.82\\
0.008 & 100& 5.52& 19.28& 19.63&  19.91& 20.18& 20.40& 20.62& 20.83\\
0.008 & 80 & 0.80& 19.23& 19.55&  19.87& 20.15& 20.39& 20.61& 20.81\\
0.008 & 70 & 0.74& 19.18& 19.51&  19.84& 20.12& 20.38& 20.59& 20.80\\
0.008 & 60 & 0.69& 18.79& 19.49&  19.80& 20.08& 20.35& 20.57& 20.79\\
0.008 & 50 & 0.68& 18.96& 19.27&  19.72& 20.02& 20.29& 20.54& 20.77\\
0.008 & 40 & 1.43& 18.90& 19.19&  19.63& 19.93& 20.23& 20.50& 20.74\\
0.004 & 200& 2.36& 19.36& 19.65&  19.94& 20.21& 20.43& 20.64& 20.85\\
0.004 & 150& 2.09& 19.26& 19.60&  19.89& 20.14& 20.38& 20.61& 20.81\\
0.004 & 120& 1.03& 19.25& 19.54&  19.86& 20.14& 20.37& 20.58& 20.79\\
0.004 & 100& 0.80& 19.23& 19.54&  19.84& 20.11& 20.35& 20.56& 20.78\\
0.004 & 80 & 0.85& 19.15& 19.45&  19.79& 20.03& 20.31& 20.55& 20.76\\
0.004 & 70 & 0.70& 19.14& 19.42&  19.73& 20.03& 20.29& 20.53& 20.75\\
0.004 & 60 & 0.55& 18.94& 19.25&  19.66& 19.97& 20.27& 20.52& 20.74\\
0.004 & 50 & 3.01& 18.82& 19.15&  19.56& 19.92& 20.23& 20.49& 20.72\\
0.001 & 200& 1.29& 19.19& 19.51&  19.81& 20.08& 20.34& 20.56& 20.78\\
0.001 & 150& 1.42& 19.13& 19.40&  19.71& 19.99& 20.25& 20.48& 20.69\\
0.001 & 120& 1.11& 18.84& 19.20&  19.62& 19.94& 20.21& 20.45& 20.66\\
0.001 & 100& 1.10& 18.99& 19.25&  19.64& 19.92& 20.19& 20.43& 20.65\\

\hline
\end{tabular}
\end{minipage}
\end{table*} 

After the end of core hydrogen burning most stars expand to become a red giant with a strong ($>10^{-4} M_{\odot} {\rm yr}^{-1}$) and slow ($\approx 100 \, {\rm km \, s}^{-1}$) wind. This produces a very dense free-wind region. If the mass-loss is not strong enough to remove the hydrogen envelope and produce a WR star, the SN occurs in this environment. The most massive stars ($\ga 80M_{\odot}$) do not experience a red giant phase and remain blue throughout their evolution. Such stars are referred to as luminous blue variables (LBVs). They have high mass-loss rates and fast winds throughout their lifetime. Such stars produce the largest bubbles as the wind injects more kinetic energy into the bubble.

If the mass loss is severe enough to remove most of, or all, hydrogen from the star a further stage of WR star evolution occurs. WR stars have strong ($\approx 10^{-5} M_{\odot} {\rm yr}^{-1}$) and fast ($\approx 1000 \, {\rm km \, s}^{-1}$) winds. This wind impacts with the slowly moving RSG wind accelerating it forming a shell travelling at a few $\times 100 {\rm km \, s^{-1}}$ \citep{garcia2,vanmarle}. If $n_{0} \ga 1 {\rm cm^{-3}}$ the free-wind region is cleared of all this red giant material, instabilities in the shell make it mix into the stalled-wind region \citep{garcia2}. If $n_{0} \la 0.1 {\rm cm^{-3}}$ this material does not penetrate deep into the stalled wind and instabilities do not have time to grow so this shell is well defined at core-collapse.

The exact position and radius separating the ISM, stalled wind and free wind varies with the initial parameters for the star and the ISM. For a given stellar mass, lower initial ISM density results in a larger bubble. More massive stars produce larger bubbles while lower metallicity stars produce smaller bubbles. Figures \ref{hydro2d} and \ref{hydro2d_B} are examples of our simulation results. In these figures we use the same stellar model but begin with different initial ISM density. As we can see the result roughly follows the simple model in figure \ref{simplefig} although the inhomogeneity of the stalled wind region can be a few orders of magnitude. We also find that the initial ISM density effects the structure of the stalled wind region. In figure \ref{hydro2d} the stalled-wind/ISM interface varies to a larger degree than that in figure \ref{hydro2d_B}. This is because with the higher density ISM the shell of RSG material swept up by the WR wind can reach the interface. Instabilities in this shell make it form into loops as it travels \citep{garcia2}. When it reaches the outer interface these loops produce the variable interface shown in figure \ref{hydro2d}. With lower ISM density the bubble is much larger and the same shell never reaches the outer interface and the instabilities have not grown to the same level as in the higher density ISM case. This leaves a shell of material moving at a few $100 \, {\rm km \, s^{-1}}$ away from the star. The strength of these instabilities maybe dependent on azimuthal resolution. If this was increased the shell may become unstable and not be as well defined as we find here.

For the GRB afterglow lightcurve however only the magnitude of the stalled-wind/free-wind interface is important since once the jet encounters this interface it will slow considerably and will take years to reach any of the large density variations at which point the afterglow will be unobservable. However these shells will have consequences for the afterglow spectrum \citep{vanmarle}, which we discuss below

We show in tables \ref{innerR} and \ref{outerR} the positions of the important interfaces in our simulations. These are the position of the free-wind to stalled-wind interface in table \ref{innerR} and the position of the stalled-wind to ISM interface in table \ref{outerR}. We calculate these values by averaging the position of these boundaries over the azimuthal direction. Looking at the tables we see how these distances vary with the three initial parameters of initial metallicity, initial mass and initial ISM density. The possible distances for $R_{\rm SW}$ range from $2 \times 10^{18}$ to $7 \times 10^{20} {\rm cm}$. The value of $R_{\rm ISM}$ varies from $10^{19}{\rm cm}$ to $9 \times 10^{20} {\rm cm}$. These distances are important when considering the ionisation of the region around the GRB progenitor. We will be able to calculate the extent of the region ionised by the prompt GRB emission and the amount of material that remains unionised leave an imprint in the afterglow absorption line spectra.

The most important parameter is the initial ISM density. Stellar parameters produce only slight variation in the bubble size. The stars that lose the most mass produce the larger bubbles as they input more energy into the bubble to make it expand further. It is interesting that the change from small bubbles to a very large free-wind regions occurs over a reasonable range of ISM densities from $10^{3}$ to $10^{-3}\, {\rm cm^{-3}}$. The model of \citet{windbubbles2} predicts that $R_{\rm ISM} \propto n_{0}^{-1/5}$ and that $R_{\rm SW} \propto n_{0}^{-3/10}$. When we compare these trends to those in tables \ref{innerR} and \ref{outerR} by extrapolating from the $n_{0}=1 {\rm cm^{-3}}$ column we find reasonable agreement. The difference between analytic and the numerical values are mostly less than 0.3 dex. The largest differences are at the highest or lowest ISM densities.

For $R_{\rm ISM}$ the main source of difference from the analytical expression is from taking the average value of $R_{\rm ISM}$. This introduces some uncertainty into where we define this interface and the effect is largest for the smallest bubble sizes. $R_{\rm SW}$ is affected by the use of non-constant wind velocities over the star's evolution. With the bubbles being different sizes the wind parameters at the interfaces change with initial ISM density. In the cases where the wind is constant in nature during the final stages of evolution the simulations reproduce the  analytic expressions well. However if the wind is changing on a timescale similar to the time it takes for the wind to reach the interface then the assumption made in the analytical models breaks down.

The density jump at the free wind/stalled wind interface is between a factor of 4 and $\approx 8$. The lower value of 4 is expected from the conditions of an adiabatic shock. Again if the wind velocity is changing on a timescale similar to the time for the wind to reach the interface the size of the jump can be increased by a small factor: an increasing wind speed will lead to a relatively less dense free-wind region and therefore a greater jump. A second reason for this increase is that in the shocked wind is also not made up of material with a uniform velocity. Therefore previously slower material that has entered the wind can slow faster material entering the shock, again increasing the jump. Most of our models only have a jump slightly higher than the pure adiabatic prediction ($<5$). The models with larger jumps are few and are confinded to the models with higher initial ISM densities (for example figure \ref{hydro2d}).

Most GRB afterglow models assume the jet traverses either a constant density medium or a free-wind medium. There are also models of afterglow lightcurves in a free-wind region followed by a constant density region \citep{dailu2002,grb030226,Enrico2}. Afterglow calculations should include the stalled-wind region. The environment can be modelled simply with the final wind parameters from stellar models used for the free-wind region. Then the beginning of stalled-wind region can be determined by assuming a radius for the density jump. The size of the jump is between a factor of 4 and $\approx 8$ than the free-wind density at that point. There are large inhomogeneities in the stalled-wind region. However these tend to be deep in the stalled wind. The afterglow jet is unlikely to propagate far once it has penetrated the stalled-wind region as it is decelerated by the density jump. Therefore when inhomogeneities are encountered the afterglow will no longer be observable.

\subsection{The Free-Wind Density}

The free-wind density is an important parameter that describes the free-wind environment. A number of authors have modelled afterglow lightcurves with free-wind profiles for the circumstellar environment. These calculations provide an estimate of the density in the region around the GRB progenitor. The values are listed in table \ref{grbdensities} taken from \citet{grbp2} and \citet{pk02}. The wind density is defined as,
\begin{equation}
\rho=\frac{A}{r^{2}},
\end{equation}
with,
\begin{equation}
A=\frac{\dot{M}}{4 \pi v_{\rm wind}},
\end{equation}
where $\dot{M}$ is the mass-loss rate and $v_{\rm wind}$ is the wind speed. However it is more convenient to use a wind density parameter, $A_{*}$, normalised to have a value of 1 for a standard WR wind,
\begin{equation}
A_{*}=\Big(\frac{\dot{M}}{10^{-5}\,{\rm M}_{\odot}\,{\rm yr^{-1}}}\Big) \Big( \frac{1000 \, {\rm km \, s^{-1}}}{v_{\rm wind}} \Big).
\end{equation}

\begin{figure}
\includegraphics[angle=0, width=84mm]{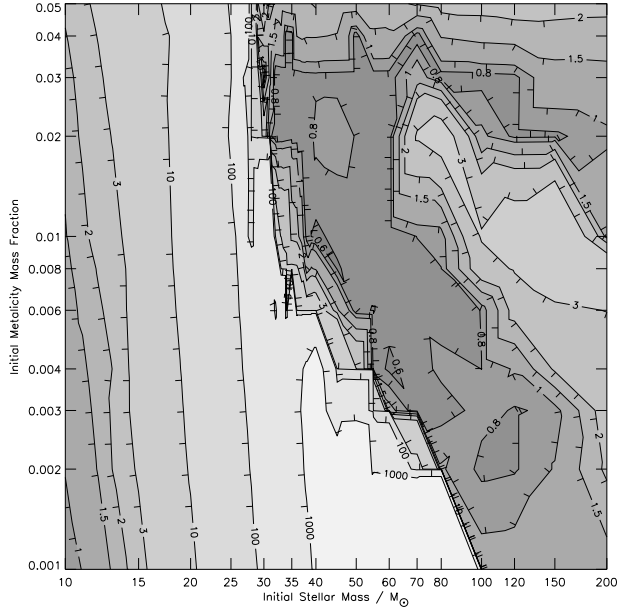}
\caption{Final $A_{*}$ value for our models including the scaling of WR wind velocities with initial metallicity. Contour values show the value of $A_{*}$ and the tick marks indicate the downhill direction.}
\label{finalA_nZ}
\end{figure}

We show in figure \ref{finalA_nZ} the $A_{*}$ values derived from our models over initial mass and metallicity. In the figure it is clear that for WR stars the $A_{*}$ value is relatively flat. For most WR stars $A_{*} \approx 0.6$ to $0.8$, the lowest value we find here is 0.55. There is a ridge in initial mass/metallicity space of WR stars with higher $A_{*}$ values up to 3. This region separates lower mass stars that have a red supergiant phase from the more massive stars that move straight from the main sequence to WR evolution undergoing an LBV phase. Therefore our full range becomes, $0.55 \la A_{*} \la 3$. Taking into consideration the uncertainty in the mass-loss rate and wind velocity equations given by \citet{NL00} the lowest $A_{*}$ value possible is 0.3. The values agree with the range of $A_{*}$ values from the WR stars in \citet{NL00} which for WN and WC stars have a range from 0.34 to a few.

This range of predicted $A_{*}$ values are of the same order, $\la 1$, as the higher inferred values in table \ref{grbdensities}. We do not find winds with $A_{*}$ similar in order to the very lowest values, $\la 0.1$, required for the first three GRBs listed in the table. The GRBs with the lower densities in the table were first thought to be better modelled by a uniform constant density environment. The low density wind solutions provide an equally good fit \citep{grbp2}. There are several possibilities on how to obtain a lower $A_{*}$ for the circumstellar environment.

First the values we present are uncertain, the WC mass-loss rates can vary by 0.15 dex with the WC wind velocities can vary by 0.13 dex. As already mentioned, the problem of clumping in the winds of hot stars is still being tackled and the uncertainties may be larger \citep{clump1,clump2}. We estimate that the $A_{*}$ values are perhaps uncertain by as much as a factor of 2 due to the mass-loss rate uncertainty.

We have so far ignored our exclusion of WO stars. The two WO stars in \citet{NL00} have $A_{*}$ values of 0.07 and 0.27 which are lower than the values for WN and WC stars. This is because WO stars have higher wind velocities than WC stars. Therefore if any of our WC models are in fact WO stars, either because our classification is wrong or they experience greater mass-loss due to rotation of binarity, then they would have a low density wind. With only a small sample of WO stars we do not know how low the $A_{*}$ value can be for these stars. The WO star values are much closer to the lower values in table \ref{grbdensities}. It could be that the wind densities in these cases could be explained by a WO progenitor rather than a WC progenitor. 

Then we can consider processes that might effect the dynamics of the circumstellar environment. The effect of rotation has been studied by \citet{rot1} and \citet{rot2}. The studies tend to indicate that the density along the stellar rotation axis, where the jet is thought to propagate, will remain constant or increase. This does not help our search for a low density wind. We have ourselves tested the effect on the wind structure if the progenitor is in a binary. The effect of binarity is discussed by \citet{Enrico2}. They conclude that the circumstellar medium will not be significantly altered in most cases. The one case that the medium may be altered will be in binaries when the masses of the two stars are close and therefore the stellar winds are of comparable strength. The outcome will be sensitive to the initial parameters of the binary and is worth of future investigation.

Another possibility is that the low $A_{*}$ values are very similar to those produced by blue supergiants (BSGs). We know it is possible for evolved BSGs to explode in a supernova. The stellar progenitor of SN1987A was a BSG. It is thought that the star underwent a merger event with a binary companion that led to a single BSG \citep{podsi92}. BSGs have similar wind velocities to WR stars but their mass-loss rates are usually 10 to 100 times lower. Therefore the typical $A_{*}$ values range from 0.1 to 0.01, agreeing with the lower values in table \ref{grbdensities}. However \citet{woosleyGRB} and \citet{limitsbyjets} rule out such progenitors due to the stars having too large a radius.

A final possibility is that the GRB progenitor is of very low initial metallicity that we do not model here ($Z/Z_{\odot}<\frac{1}{20}$). Such stars will only become WR stars in binary systems due to their feeble stellar winds \citep{KD2002}. Although recent models by \citet{langeryoon} suggest the stars might undergo homogeneous evolution during hydrogen burning.

\begin{table}
\caption{Value of $A_{*}$ for various GRBs, adapted from \citet{grbp2} and \citet{pk02}.}
\label{grbdensities}
\begin{tabular}{lc}
\hline
GRB	&	$A_{*}$\\
\hline
011121	&	0.02\\
020405	&	$<0.07$\\
021211	&	$\sim$0.015\\
\hline
970508	&	0.3,0.39\\
991208	&	0.4,0.65\\
991216	&	$\sim$1\\
000301C	&	0.45\\
000418	&	0.69\\
021004	&	0.6\\
\hline
\end{tabular}
\end{table} 

\section{Modelling the GRB afterglow lightcurve.}

After the prompt emission of gamma-rays in the GRB itself the relativistic jet continues its motion and is decelerated by the material around the progenitor causing a shock that transfers energy from the jet to radiation that is emitted. This is the GRB afterglow. The emission slowly decays with time usually as a smooth power law. There are some cases however where the afterglow is far from smooth and in some cases rebrightens considerably before continuing to decay. In this section we will not attempt to fit model lightcurves to observed GRB afterglows but will instead show the relative effects of varying the circumstellar environment density profile.

In our calculations we assume that the initial kinetic energy in the relativistic jet has an isotropic equivalent value, $E_{\rm iso}=10^{53}$ ergs. This corresponds to a true energy, $E=(1-\cos \theta)E_{\rm iso}=4 \times 10^{50}$ -- $3 \times 10^{51}$ ergs, for jet opening angles in the range 5 -- 15$^{\circ}$, in agreement with observations \citep{frailjet,pkjet}.

We follow the dynamical evolution of the relativistic jet using the simple formalism for energy and momentum conservation derived by \citet{grbagcalc}. The calculation is stopped just before the jet enters the non-relativistic regime (in practice at $v=0.75c$). Once the evolution of the Lorentz factor is known, the synchrotron emission of the jet is computed following \citet{grbafterglowspec}. It is assumed that a fraction, $\epsilon_{\rm B}$ of the internal energy in the post-shock region is transferred into the magnetic field and that a fraction, $\epsilon_{\rm e}$, is injected into a population of non-thermal relativistic electrons, with a power-law distribution of slope, $p$. We have adopted, $\epsilon_{\rm B}=10^{-3}$, $\epsilon_{\rm e} = 0.1$ and $p=2.5$, which are typical values found in afterglow fits. Finally the effect of the opening angle of the jet and the viewing angle of the observer is included following the formula given in \citet{openningangles}. Here, we have focused on on-axis observations (the line of sight and the jet axis are the same) but we have considered different opening angles of the jet: $\theta=5$, 10 and 15$^{\circ}$ as well as a spherical ejecta for comparison (isotropic ejecta). We then calculate the resulting flux in the V band, with the GRB at a redshift of one.

\begin{figure}
\includegraphics[angle=0, width=84mm]{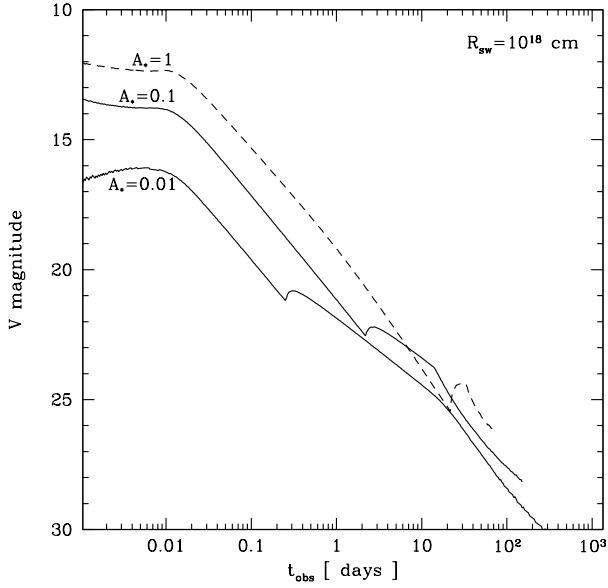}
\caption{Examples of afterglow lightcurves with varying free-wind density. $A_{*}=0.01$, 0.1 and 1.0 while $R_{\rm SW}=10^{18}{\rm cm}$. The opening angle is constant at 10$^\circ$. The dashed line has been used to avoid confusion in the plot at late times.}
\label{lc1}
\end{figure}

\begin{figure}
\includegraphics[angle=0, width=84mm]{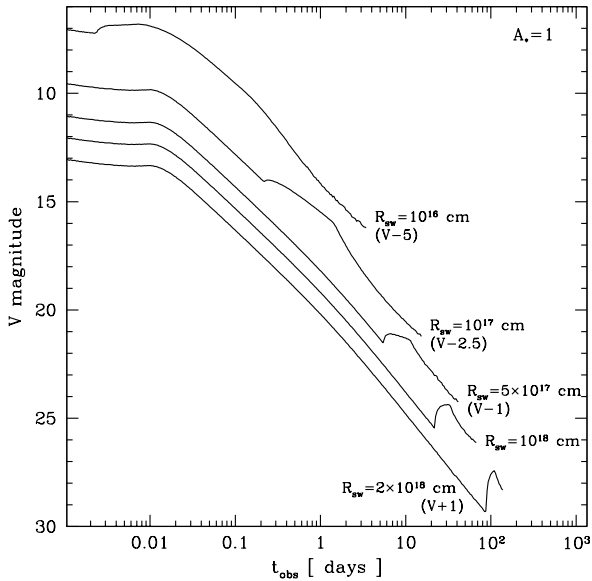}
\caption{Similar to figure \ref{lc1} but with $A_{*}=1$ and various values for $R_{\rm SW}$. The opening angle is constant at 10$^\circ$.}
\label{lc2}
\end{figure}

\begin{figure}
\includegraphics[angle=0, width=84mm]{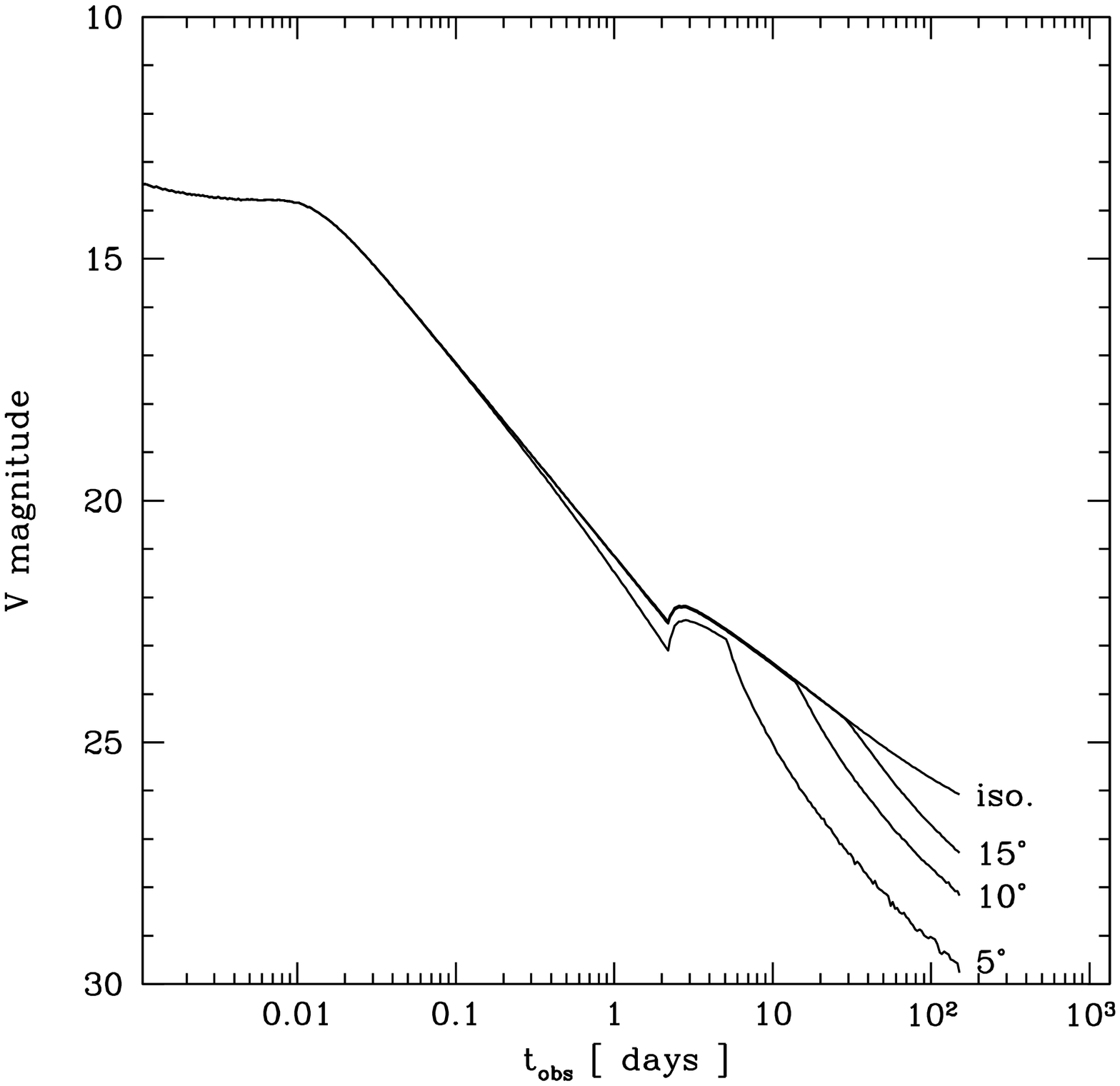}
\caption{Similar to figure \ref{lc1} but with  $A_{*}=0.1$ and $R_{\rm SW}=10^{18}{\rm cm}$. We vary the opening angle.}
\label{lc3}
\end{figure}

In figures \ref{lc1}, \ref{lc2} and \ref{lc3} we compare results from calculations with different values of the wind density parameter $A_*$, boundary radius between the free wind and stalled wind regions, $R_{\rm SW}$, and the jet opening angle, $\theta$. The light curves are characterised by an initial slope corresponding to a wind environment, a jump when the shock reaches $R_{\rm sw}$ followed by a shallower slope in the constant density medium. For some of the lightcurves the break expected when $1/\Gamma=\theta$ is observed after the jump. 

Figure \ref{lc1} demonstrates the effect of varying $A_{*}$ while keeping $R_{\rm SW}$ constant at $10^{18}{\rm cm}$. A denser wind has a more intense afterglow and is decelerated more quickly. Therefore the jet takes a longer time to reach the density jump at the free/stalled wind interface. The slight bump in the lightcurves occurs when the jet encounters this interface. In the case when $A_{*}=1$ the bump is observed the latest as the dense wind decelerates the jet rapidly. For a very dense wind ($A_{*} \approx 10$) the jet can be decelerated very quickly so that the reverse shock is efficient during the GRB. This may affect the prompt GRB emission if the emission mechanism are internal shocks in the jet.

In figure \ref{lc2} we show the differences introduced by moving the position of the interface in our density profiles. For the smallest distances the effect of the jump is hidden in the general shape of the light curve with only pronounced bumps at later times. This figure also demonstrates the different slopes of the lightcurves decay when the jet propagates through a constant density or free-wind medium. Importantly when $A_{*}$ and $R_{\rm SW}$ are set to the values in table \ref{innerR} we find that the rebrightening would occur after 10 days in most cases. At this time the afterglow is already dim and such effects maybe difficult to observe.

Finally figure \ref{lc3} shows the effect of altering the opening angle. The possible afterglow lightcurves are interesting. All follow the same pattern and all have substantial brightening at a few days after the GRB due to the jet encountering the free-wind/stalled-wind interface. The greatest effect on the lightcurve comes from the break as the jet is decelerated and $1/ \Gamma$ drops below the opening angle of the jet.

The most important value to fit with any calculations is the initial value of $A_{*}$. This will provide a solid constraint on the progenitor as it is directly related to the final mass loss of the massive star. The position of the jump can be found from rebrightening in the afterglow but is of limited use because it is affected not only by the details of the progenitor but also by the initial density of the ISM and the uncertainty in the wind velocity of pre-WR stars. Although in the cases where the afterglow is better described by a constant density medium this can be explained by $R_{\rm SW}$ being smaller than around $10^{16} {\rm cm}$ so any signature of the wind profile will be in the very early afterglow lightcurve.

\section{Absorption lines in GRB afterglow spectra}

The first GRB afterglow was observed for GRB970228 \citep{firstgrbag}. The wait for a redshift from a GRB afterglow was short, it was observed for the second observed afterglow of GRB970508 \citep{firsggrbZ}. From a spectrum the redshift of a GRB can be determined from lines intrinsic to the host galaxy or absorption lines within the host galaxy. In some cases there are extra absorption lines that have a small difference to the redshift of the host galaxy and the GRB. The inferred velocities of this material are a few hundred and a few thousand kilometres per second. Upon discovering these lines it has been suggested that they might be caused by absorption by the GRB progenitors stellar wind material \citep{first4wrwinds,grb1,grb2,grb3}. Although there are alternative explanations such as structures related to the host galaxy rather than the GRB progenitor. 

With our stellar models and simulated circumstellar environment we have information on the velocity profiles we can expect from the stellar wind. The predicted velocities for the free-wind region do match well with the velocities of the absorption lines at the highest velocity offset. In the stalled-wind region we also predict velocities of a similar magnitude but we find the exact figure depends quite sensitively on the initial ISM density and the wind velocities during the main-sequence and red-supergiant phases of evolution. This agreement indicates stellar wind material could be responsible for these absorption lines.

The first main question is whether the ions observed are present in the pre-GRB wind. We verified this by modelling the environment using the program Cloudy V96.01 \citep{cloudy}. We used a black body spectrum for the WR star emission at effective temperature and luminosity of WR stars. We found that the observed species are present in the free-wind region.

The second, and more important, question is whether the GRB completely ionises the immediate environment as discussed by \citet{grbsurf} and \citet{grb3}. Therefore only fully ionised species will be present which cannot produce absorption lines. In \citet{grbsurf} the comment is justified by calculating the recombination time-scale of the material and finding it was much longer than the time the afterglow is observed for. While in \citet{grb3} they estimate this in terms whether there is enough energy at a certain radius to complete ionise the material.

We also investigate if the initial emission would be able to completely ionise the surrounding medium. We start by considering the ionisation cross-section, $\sigma(E)$, where $E$ is the energy of the incident photon. We approximate the cross-sections energy dependence as,
\begin{equation}
\sigma(E) \approx \sigma_{\rm i} \big(\frac{E}{E_{\rm i}} \big)^{-3},
\end{equation}
when $E_{\rm i}$ is the minimum energy required to ionise the species in question. Our next step is to estimate the number of ionisations per atom or ion at a distance $R$ from the GRB,
\begin{equation}
n_{\rm i}(R) = \frac{1}{4 \pi R^{2}}{ \int_{E_{\rm i}}^{\infty} \sigma(E) N(E) dE},
\label{cooleqn}
\end{equation}
where $N(E)$ describes the spectrum of the GRB such that $N(E) dE$ is the number of photons produced by the source between $E$ and $E+dE$. We also define $R_{\rm max}$ such that $n_{\rm i} (R_{\rm max}) =1$ gives the maximum distance where the atom or ion can be ionised. Before we evaluate this in integral we must determine the GRB spectrum. We approximate the GRB spectrum to be,
\begin{eqnarray}
N(E)  =  \zeta \Big(\frac{E}{E_{\rm p}} \Big)^{\alpha} & = \zeta \big(\frac{E}{E_{\rm p}} \big)^{-1} & E<E_{\rm p} \\
  & = \zeta \big(\frac{E}{E_{\rm p}} \big)^{-2.5} & E>E_{\rm p}. \\
\end{eqnarray}
Where $E_{\rm p}$ is the peak energy of the burst spectrum. $\zeta$ is a constant found by normalising with the isotropic radiated energy of the burst, $\zeta = E_{\rm iso}/(2 E_{\rm p}^{2})$. With these assembled formulae we evaluate equation \ref{cooleqn}. The result is,
\begin{equation}
n_{\rm i}(R) =\frac{ \sigma_{\rm i} E_{\rm iso} }{24 E_{\rm p} \pi R^{2}}.
\end{equation}

We take the ion C(IV) as our example as this has been observed in some GRB afterglow spectra. We use the burst parameters from GRB021004 which we discuss later. Therefore we have $ \sigma_{\rm i}= 0.66 \times 10^{-19} \,{\rm cm^{2}}$, $E_{\rm iso}=2 \times 10^{52}\, {\rm ergs}$, $E_{\rm i}=46\, {\rm eV}$ and the rest frame $E_{\rm p} = 200\,{\rm keV}$. Using these values we find $R_{\rm max} = 2.3 \times 10^{19}\, {\rm cm}$. From this we can assume that nearly all C(IV) in the free-wind  material is ionised, this argument despite being very simplistic means that absorption lines in the afterglow spectra are probably from material not associated with the GRB progenitor. But we only look at one species of one element, the combination of all elements present may increase the absorbing power of the circumstellar material. We can also find that some of the free-wind regions in our simulations are larger than this size if we start with a low ISM density. At initial ISM densities of $n_{0} \le 1 {\rm cm^{-3}}$, the $R_{\rm SW}$ is greater than this ionisation radius so it is possible for some material to remain unionised and produce absorption lines in the afterglow spectra. Comparing to the ionisation radii calculated in \citet{grb3} we can make a similar conclusion.

There are also other details to consider. \citet{grb3} suggested the jet may be structured. The jet may be formed from a central hyper-relativistic core surrounded by less relativistic material that gives rise to the afterglow. There is a similar alternative explanation that uses the temporal structure to the jet. Initially the jet will be tightly confined as will its emission by the large beaming factor due to the relativistic motion. This means all the ionising flux will be tightly beamed through only a small fraction of material directly in its path. However after the initial burst and the fireball propagates it will spread out. Therefore while the material down the centre of the jet will be ionised the material at the edge of the jet will not be. This becomes more true the larger the angle between the jet axis and the line of sight.

\begin{figure}
\includegraphics[angle=0, width=84mm]{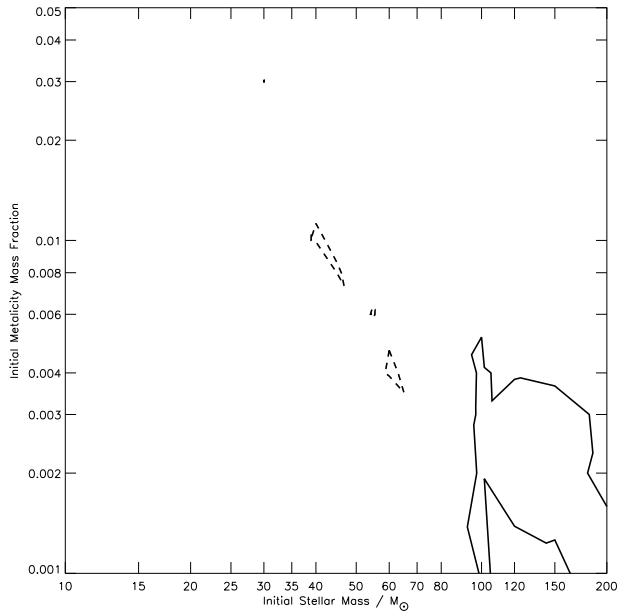}
\caption{The solid line show where $v_{\rm wind}=3000 {\rm km \, s^{-1}}$. The dashed line shows where $A_{*}=0.6$.}
\label{grb021004a}
\end{figure}

\begin{figure}
\includegraphics[angle=0, width=84mm]{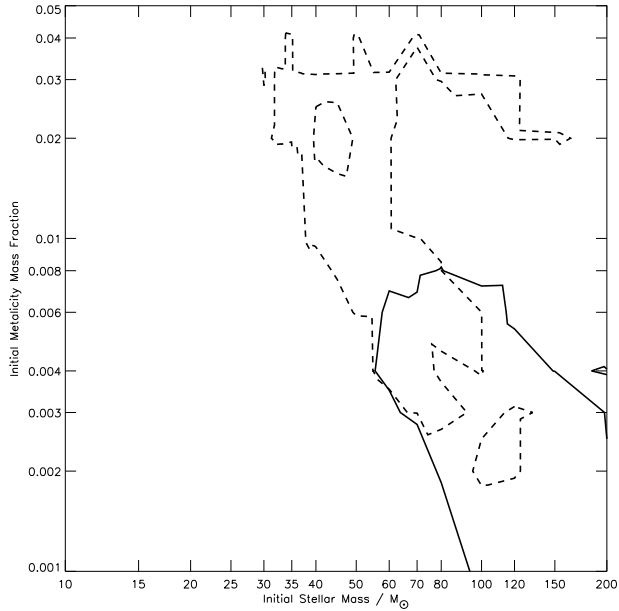}
\caption{Same as figure \ref{grb021004a} but the wind velocity from our models has been increased by 0.13 dex, the uncertainty of the predicted wind velocities. The solid line show where $v_{\rm wind}=3000 {\rm km \, s^{-1}}$. The dashed line indicated the region where $A_{*}=0.6$.}
\label{grb021004b}
\end{figure}

We conclude that the lines we see could come from the GRB progenitor's wind in some cases. Only observations of more GRB afterglow spectra will indicate if this is a general trend. We also predict that there will be an evolution of the observed spectra with time. For example eventually the jet will reach beyond the free wind, at that point the higher velocity absorption lines should disappear from the spectra. If they were to remain then they must be external in origin. Although it is very improbable the afterglow will remain luminous enough at late times to allow detailed spectroscopy.

\section{Comparing to Observed GRBs}

While there are a large number of observed GRBs and a good fraction with observed afterglows there are relatively few with observed absorption line spectra. Below we discuss three GRBs that have observed afterglows and compare them with our predicted circumstellar environments.

\subsection{GRB021004}
The position of this burst was localised within a few minutes by HETE-2. This meant that detailed optical afterglow observations were taken. Many different groups took spectra of the afterglow \citep{first4wrwinds,grb1,grb2,grb3}.

\citet{grb3} found there were absorption lines for C(IV), Si(IV) and Hydrogen, blueshifted with respect to the host galaxy with velocities of around $3000$ and $500 \, {\rm km \, s^{-1}}$. Their interpretation was that these lines were caused by material in the stellar wind of the GRB progenitor. 

Possible resolutions of the ionisation problem were discussed above. However the presence of hydrogen should not be possible if the progenitor star was a WR star. \citet{grb3} suggested the WR star was in a binary with a main-sequence O star companion. The O star wind would mix hydrogen into the WR wind. Both stars have similar wind velocities and the majority of the material in the resultant wind would still originate from the WR star.

If we assume that the absorption lines are from material in the stellar wind we can ask, `what does the afterglow and absorption lines tell us about the progenitor itself?' \citet{grbp2} estimate an $A_{*}$ value of 0.6 for this GRB. On figure \ref{grb021004a} we plot the region where $A_{*}=0.6$ in initial mass and metallicity space. On the same plot we include contours of the free-wind speed determined from the absorption lines of $v_{\rm wind}=3000 {\rm km\, s^{-1}}$. These contours do not overlap. However the mass-loss rates and the wind velocities are uncertain. In figure \ref{grb021004b} we increase all predicted wind velocities by 0.13 dex, the quoted uncertainty in the wind velocity equations, an overlap now occurs. The inferred free-wind density and wind speed then agrees with the predicted details of a low metallicity WC star. It must be noted that we are only able to achieve an agreement with the observations by exploiting that the winds are uncertain. Until the uncertainties in WR stars and their winds are greatly reduced we cannot make an identification in this way.

An alternative explanation is that the progenitor may be a WO star. This is possible as while WO stars tend to have higher wind velocities and a lower free-wind density, the progenitor could have been a transition object with faster wind speeds than those predicted that our solar metallicity WC stars. Although there are very few observed WO stars so we do not fully understand the possible wind speeds. Further more we have not considered the uncertainty in the WR mass-loss rates. Altering the magnitude of the WR mass-loss rates will effect our models and change the masses, surface composition and luminosity of our models and therefore change the final velocities and mass-loss rates. Combining all the possible uncertainties it is not possible to make any conclusions about the progenitor of GRB021004 other than its observed details agree with a wide range of WR models.

The low velocity lines provide extra information on the wind structure around the progenitor. \citet{vanmarle} discuss how the lines are likely to be caused by the shell formed when the slower RSG wind material is swept up and accelerated by the faster WR wind. We find this occurs in all our simulations were the stars go through a RSG phase. For initial ISM densities, $n_{0} \ga 1 {\rm cm^{3}}$, this shell reaches the stalled-wind region, decelerates and mixes into the stalled wind. At low initial densities ($n_{0} \la 0.1 {\rm cm^{3}}$) we find that the shell does not reach the stalled-wind region and remains in the free-wind region. The velocity is typically a few $\times 100 {\rm km \, s^{-1}}$, for a number of our WR stars the shell is close to the observed $500 \, {\rm km \, s^{-1}}$. Therefore the lower velocity lines may give a qualitative indication of the initial ISM density. Although the structure of the low velocity lines is complex and made up of many different single lines over a range of a few $\times 100 {\rm km \, s^{-1}}$. Therefore no simple interpretation is possible.

In summary the progenitor may have been an WR star at any metallicity. However such stars above solar metallicity will lose too much mass to form a black hole directly at core-collapse, therefore sub-solar stars are favoured (figure \ref{remnant}. 

\subsection{GRB020813, GRB030226 and other bursts.}

There are two more GRBs with absorption lines slightly offset from the host galaxy redshift. For GRB020813 the inferred wind velocity was 4320 km s$^{-1}$ with an $A_{*} \approx 0.01$ \citep{grb020813}. For this GRB the case of a WO star is slightly stronger as the wind velocity and wind density agree with the typical values for observed WO stars.

For GRB030226 the velocity was 2300 km s$^{-1}$ \citep{226a,226b,226c}. There an $A_{*}$ has not been estimated. However the wind velocity is too low for a WO star and is more in agreement with the wind speed of a WC star.

\section{Discussion}

The study of GRBs and their afterglows is a rapidly expanding field; it is not only of interest to those who wish to understand GRBs but also those who wish to use GRBs as tools to solve other problems, such as investigating star-formation history or the Lyman-$\alpha$ forest. In this paper we have both described the circumstellar environment of possible GRB progenitors and discussed the effect of such an environment on the GRB afterglow. 

The details of the progenitor that affect a GRB afterglow are the final wind velocity and mass-loss rate, these together determine the $A_{*}$ value of the free-wind region around the star. This value affects the slope of the afterglow lightcurve. For some environments the afterglow jet might encounter the density jump at the interface between the free and stalled wind while it remains observable \citep{dailu2002,grb030226,Enrico2}. This density jump is between a factor of 4 and $\approx 8$. It is easy to use a simple model density profiles of the free-wind region connected to a stalled-wind region. The circumstellar environment in current afterglow calculations only considers the free-wind profiles and ignore the stalled-wind region even though it has been known to exist for sometime \citep{windbubbles1}. The environment can be described by $A_{*}$, $R_{\rm SW}$ and $\Delta \rho_{\rm jump}$. From varying these three numbers it should be possible to reproduce afterglow lightcurves, using the values of $A_{*}$ and $R_{\rm SW}$ from our stellar models as listed in table \ref{innerR}, with $4 \le \Delta \rho_{\rm jump} \la 8$. 

$A_{*}$ have been inferred from observations of GRB afterglow lightcurves. The higher values ($A_{*} \ga 0.5$) are similar to those from stellar evolution models of WC stars. Further more, using the inferred free-wind velocity from GRB afterglows we have demonstrated how it may be possible to estimate the range of possible initial parameters for the progenitor. No firm conclusions can be drawn as the uncertainty in WR wind velocities and mass-loss rates are considerable. As our understanding of these objects grows it may become possible to use the method presented here to limit the parameter space of the progenitor. One important question that must be answered is how the mass-loss rate and wind velocity scales with initial metallicity. For radiatively driven winds it is known that separate elements are responsible for determining the mass-loss rate and wind velocity. The many but weak lines of the iron group elements determine the mass-loss rate while the few but strong lines of the CNO elements determine the wind velocity \citep{VKL2001}. The iron group elements are not depleted on the surface over the lifetime of a star but the CNO elements become enhanced as time passes. For WR stars the CNO abundance is substantially different to the initial metallicity. The dependence on initial metallicity may therefore be very weak.

Of further interest will be the effect of rotation and/or a binary companion on the progenitor star and the circumstellar environment. Rotating models of GRB progenitors do retain enough angular momentum to form an accretion disk around the forming black hole to produce a GRB \citep{hmm2005}. Magnetic fields remove this possibility \citep{langerthingy}. But binary stars can spin-up a star by mass-transfer events therefore they may be the more likely progenitors \citep{langerthingy,izzy}.

The circumstellar environment is the next step to consider, it is not clear how rotation will change the details of the circumstellar free-wind region \citep{rot1,rot2}. \citet{Enrico2} present a simulation of the circumstellar environment for a rotating single star. They find that both the free-wind density and $R_{\rm SW}$ are strongly dependent on latitude.

A similar effect can be expected for binary stars as interactions can spin up the stars to rotate more rapidly than might be expected for single stars. For binary stars, because of the many possible evolutionary paths, there are a number of other possible effects on the environment. For a wide binary interactions may shorten or remove the RSG phase by enhancing mass-loss therefore there will only be a brief phase when the wind speed drops below $1000 {\rm km \, s^{-1}}$. For closer binaries common-envelope evolution may occur leading to a tight WR/O star binary where the mass-loss rate of the WR star maybe enhanced by the companion leading to larger numbers of WO stars. The most important binary cases will be those where the stellar wind of the secondary is of similar strength to the WR stars when the wind density maybe be effected \citep{Enrico2}. 

\section{Conclusion}
\begin{itemize}
\item Our predicted value of $A_{*}$ for single WC stars span a small range, $0.55  \la A_{*} \la 3$. Due to the uncertainty in WC star winds the lower value may be as low as 0.3.
\item All our WR models end as WC stars and we have no WO stars. WO stars have higher wind velocities than WC stars and therefore may give rise to the lower $A_{*}$ than we predict from out models.
\item The typical observed free-wind velocities inferred from absorption line spectra of GRB afterglows agree with predictions of pre-SN WR wind speeds from single non-rotating stars.
\item It is possible for the free-wind region to become extremely large, in some cases larger than the ionisation radius from the prompt GRB emission. Therefore it is possible for the wind material to remain unionised and be observed in the afterglow absorption line spectrum if the initial ISM density is $n_{0} \le 1 {\rm cm^{-3}}$.
\item Rapid changes in the wind velocity during the evolution of stars can produce small deviations in the expected dimensions of the circumstellar environment from analytic models.
\item We indicate that from limits on the value of $A_{*}$ and the final wind speed from the GRB progenitor it may in future be possible to estimate the initial parameters of GRB progenitors. Currently the uncertainties in WR mass-loss are too great to draw any such conclusions.
\end{itemize}

\section{Acknowledgements}
JJE would like to thank Chris Tout, Pierre Lesaffre and Richard Alexander for fluid conversations and Jorick Vink for windy conversations. We would also like to thank the referee's Norbert Langer and Allard-Jan van Marle for their many useful questions that led to an greatly improved article. We also wish to thank Mike Barlow for some details of WO stars.

\bsp

\end{document}